\documentclass[twocolumn]{aastex62}
\usepackage{mathtools}          

\usepackage{float}

\graphicspath{{./}{figures/}}

\shorttitle{Detecting Interstellar Objects Through Stellar Occultations}
\shortauthors{Siraj \& Loeb}

\begin{document}

\title{Detecting Interstellar Objects Through Stellar Occultations}

\email{amir.siraj@cfa.harvard.edu, aloeb@cfa.harvard.edu}

\author{Amir Siraj}
\affil{Department of Astronomy, Harvard University, 60 Garden Street, Cambridge, MA 02138, USA}

\author{Abraham Loeb}
\affiliation{Department of Astronomy, Harvard University, 60 Garden Street, Cambridge, MA 02138, USA}

\keywords{asteroids: individual (A/2017 U1)}



\begin{abstract}
Stellar occultations have been used to search for Kuiper Belt and Oort Cloud objects. We propose a search for interstellar objects based on the characteristic durations ($\sim 0.1 \mathrm{s}$) of their stellar occultation signals and high inclination relative to the ecliptic plane. An all-sky monitoring program of all $\sim 7 \times 10^6$ stars with $R \lesssim 12.5$ using 1-m telescopes with $0.1 \; \mathrm{s}$ cadences is predicted to discover $\sim 1$ interstellar object per year.

\end{abstract}

\keywords{Minor planets, asteroids: general -- meteorites, meteors, meteoroids}


\section{Introduction}

`Oumuamua was the first interstellar object (ISO) reported in the Solar System \citep{Meech2017, Micheli2018}. Follow-up studies of `Oumuamua were conducted to better understand its origin and composition \citep{Bannister2017, Gaidos2017, Jewitt2017, Mamajek2017, Ye2017, Bolin2017, Fitzsimmons2018, Trilling2018, Bialy2018, Hoang2018, Siraj2019a, Siraj2019b, Seligman2019, 2019arXiv190108704S}. `Oumuamua's size was estimated to be $\lesssim 200 \; \mathrm{m}$, based on Spitzer Space Telescope constraints on its infrared emission given its expected surface temperature based on its orbit \citep{Trilling2018}.

In addition to `Oumuamua, CNEOS 2014-01-08 \citep{Siraj2019a} was tentatively the first interstellar meteor discovered larger than dust, and 2I/Borisov \citep{Guzik2019} was the first confirmed interstellar comet.

Transient diffraction patterns caused by the occultation of a distant star due to an intervening small body have been proposed and used to search for Kuiper Belt objects (KBOs) and Oort Cloud objects (OCOs) in the Solar System \citep{Bailey1976, Dyson1992, Roques2000, Nihei2007, Schlichting2009, Schlichting2012, Arimatsu2019}. Here, we propose an analagous search for interstellar objects (ISOs), flagged by their unusual inclinations and unique kinematics leading a distribution of characteristic distribution of durations, the peak of which lies between that of Kuiper belt objects and that of Oort cloud objects.

\begin{figure*}[!th]
  \centering
  \includegraphics[width=.7\linewidth]{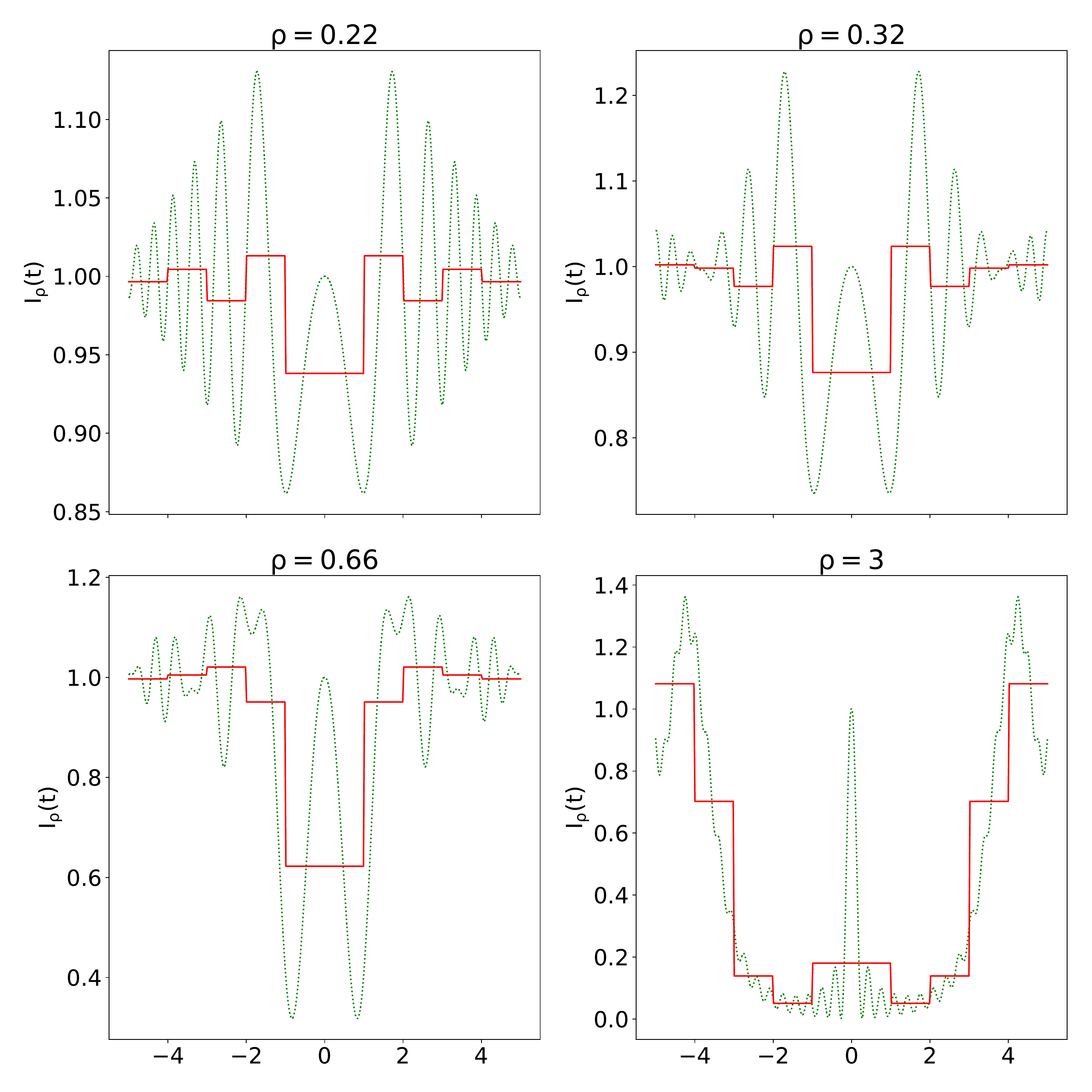}
    \caption{Occultation lightcurves for the values of $\rho_{min}$ (ratio of ISO radius to the Fresnel scale) in Table \ref{tab:1} (dotted green lines), and sampled lightcurves at temporal resolution, $t_F$ (solid red lines). For a transverse speed $v_{\perp}$ at a distance $D_L$, the time unit, $t_F = 0.1 \mathrm{s \;} (D_L / 20 \mathrm{\;AU})^{1/2} (v_{\perp} / \mathrm{10 \; km \; s^{-1}})^{-1}$.}
    \label{fig:fig1}
\end{figure*}

The data from such a search would calibrate population parameters for ISOs crucial for constraining formation theories of exoplanetary systems \citep{Duncan1987, Charnoz2003, Veras2011, Veras2014, Pfalzner2015, Do2018, Raymond2018,  Hands2019, Pfalzner2019, Siraj2019d}. Below we derive the physical characteristics and rate of the expected ISO occultation events.

\section{Theory}
\label{sec:methods}

The measured intensity at wavelength $\lambda$ from a diffraction pattern created by a spherical object of radius $R_L$ at a distance $D_L$ is,

\begin{equation}
    I_{\rho}(t) =
  \begin{cases}
    U_0^2 (\rho, t / t_{F}) + U_1^2 (\rho, t / t_{F}), & \text{$t / t_{F} \leq \rho,$} \\
    1 + U_1^2 (\rho, t / t_{F}) + U_2^2 (\rho, t / t_{F}) \\
    - 2 U_1 (\rho, t / t_{F}) \sin \frac{\pi}{2} (\rho^2 + (t / t_{F})^2) \\
    + 2 U_2 (\rho, t / t_{F}) \cos \frac{\pi}{2} (\rho^2 + (t / t_{F})^2), & \text{$t / t_{F} \geq \rho,$} \\
  \end{cases}
\end{equation}
where $\rho = (R_L/F)$ is the radius of the object in units of the Fresnel scale $F = \sqrt{\lambda D_L / 2}$, and $t_F$ is the Fresnel scale crossing time for the object \citep{Nihei2007, Roques1987}. For a transverse speed, $v_\perp$, at a distance $D_L$, the Fresnel time is $t_F = 0.12 \mathrm{s \;} (D_L / 20 \mathrm{\;AU})^{1/2} (v_{\perp} / \mathrm{10 \; km \; s^{-1}})^{-1} (\lambda / \mathrm{\mu m})^{1/2}$ and the Fresnel scale is $F = 1.2 \mathrm{km \;} (D_L / 20 \mathrm{\;AU})^{1/2} (\lambda / \mathrm{\mu m})^{1/2}$. The Lommel functions are defined as,
\begin{equation}
    U_n (\mu, \nu) = \sum^{\infty}_{k = 0} (-1)^k \left(\frac{\mu}{\nu}\right)^{n + 2k} J_{n + 2k} (\pi \mu \nu) \; \;,
\end{equation}
where $J_n$ is a Bessel function of order n. With regard to the impact parameter $b$, we assume $b = 0$ for simplicity. Figure \ref{fig:fig1} shows lightcurves for different values of $\rho$, as well as lightcurves sampled as at a spatial resolution of $F$. While the calculated lightcurves represent monochromatic light, the finite filters used in practice slightly reduces the statistical power.

For a solar-type star with an R-magnitude of 12, the flux of R-band photons at Earth is $\sim 2.8 \times 10^5 \mathrm{\; m^{-2} \; s^{-1}}$. For an intensity dip caused by an occultation event, a signal-to-noise ratio of $\gtrsim 10$ for a telescope with $1 \; \mathrm{m}^2$ collecting area with a temporal resolution of $t_f$ would require the intensity dip to be $\gtrsim 6 \%$, corresponding to $\rho \gtrsim 0.22$. The appropriate values of $\rho_{min}$ as a function of magnitude are listed in Table \ref{tab:1}.

\begin{figure*}[!th]
  \centering
  \includegraphics[width=.5\linewidth]{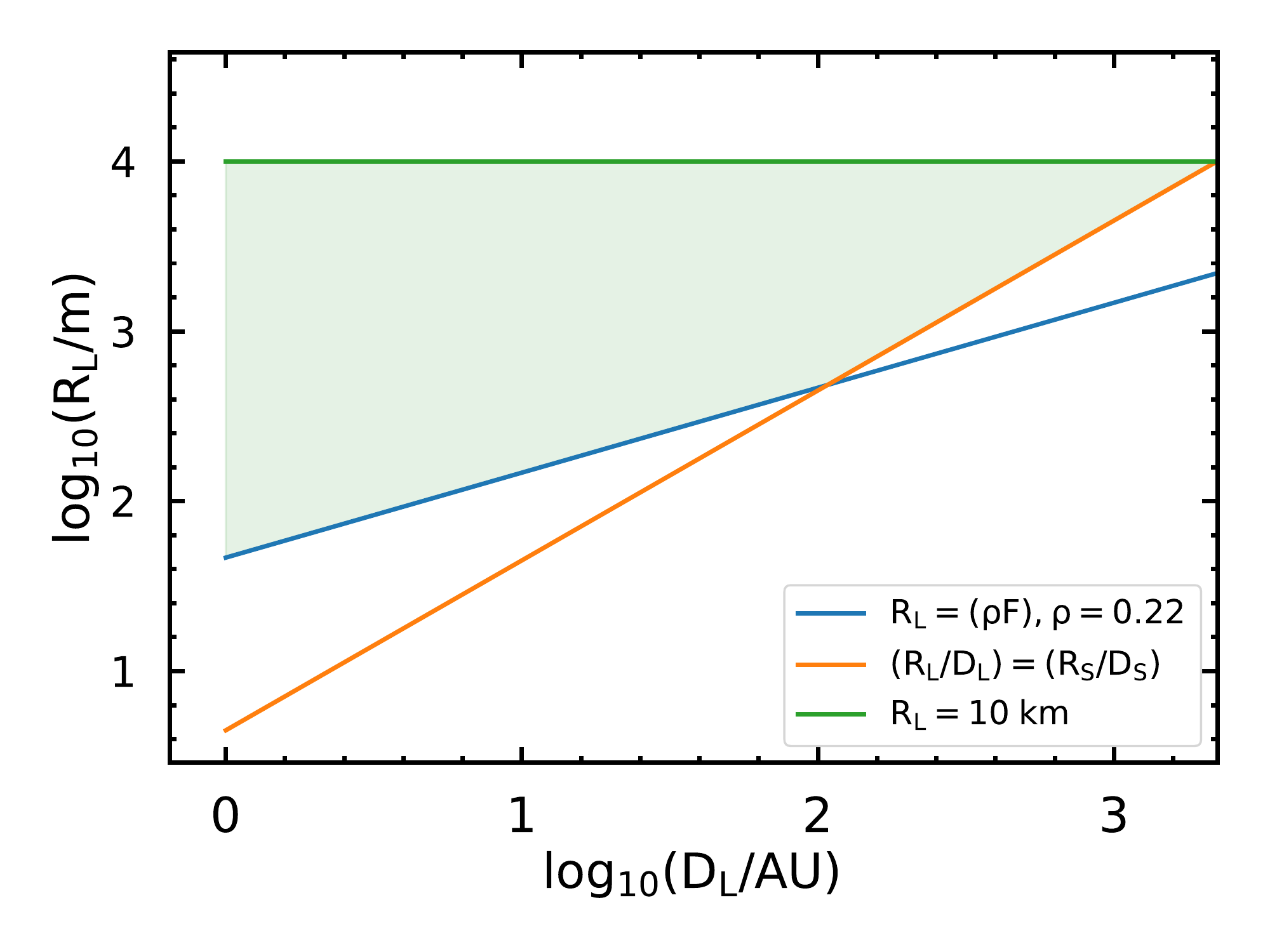}
    \caption{Parameter space in which conditions (\ref{eq:2}), (\ref{eq:3}), and (\ref{eq:4}), are satisfied for a R = 12 solar-type star. $R_S \sim R_\odot$ and $D_S \sim 300 \; \mathrm{pc}$ for an R = 12 solar-type star.}
    \label{fig:fig2}
\end{figure*}

In addition to $\rho \geq \rho_{min}$ for a strong signal, we also require the occulting object to subtend a larger angular size on the sky than the star to avoid dilution of the signal by the finite source size. We adopt the conservative condition $R_{L} \leq 10 \mathrm{\; km}$ for ISOs. Given a star of radius $R_S$ and distance $D_L$, the three conditions for an ISO occultation event are expressed as follows:

\begin{equation}
\label{eq:2}
    R_L > \rho_{min} \sqrt{\frac{\lambda D_L}{2}} \; \; ,
\end{equation}

\begin{equation}
\label{eq:3}
    \frac{R_L}{D_L} > \frac{R_S}{D_S} \; \; ,
\end{equation}

\begin{equation}
\label{eq:4}
    R_{L} < 10 \mathrm{\; km} \; \; .
\end{equation}
Figure \ref{fig:fig2} illustrates a parameter space in which all three conditions are satisfied.

We approximate the cumulative number density of ISOs as a power law with exponent -3.3, calibrated by the number density ($n_O \sim 0.2 \; \mathrm{AU^{-3}}$) of `Oumuamua size ($R_O \sim 100$ m) objects \citep{Siraj2019c, Do2018, Landgraf2000}.

\begin{table}[!th]
	\centering
	\caption{Background photon flux, maximum dip in intensity, and minimum value of $\rho$ for occultation events with solar-type background stars, assuming a $1 \; \mathrm{m}^2$ collecting area and a temporal resolution of $t_f$.}
	\label{tab:1}
	\resizebox{\linewidth}{!}{\begin{tabular}{cccc}
        \hline
        \textbf{R Magnitude} & \textbf{R-Band Photon Flux ($m^{-2} s^{-1}$)} & \textbf{$\Delta I_{min}$ ($\%$)} & \textbf{$\rho_{min}$} \\
        \hline
        12           & $2.8 \times 10^5$           & 6        & 0.22          \\
        14          & $4.4 \times 10^4$            & 15        & 0.32          \\
        16         & $6.9 \times 10^3$            & 38        & 0.66          \\
        18         & $1.1 \times 10^3$          & 95        & 3          \\
        \hline
    \end{tabular}}
\end{table}

We adopt the three--dimensional velocity dispersion of stars in the thin disk of the Milky Way as a proxy for the kinematics of ISOs, each corresponding to the standard deviation of a Gaussian distribution about the local standard of rest (LSR): $\sigma_x = 35 \mathrm{\;km\;s^{-1}}$, $\sigma_y = 25 \mathrm{\;km\;s^{-1}}$, $\sigma_z = 25 \mathrm{\;km\;s^{-1}}$ \citep{Bland-Hawthorn2016}. The resulting distribution of observed transit speeds (once the motion of the Earth is subtracted) has a mean value of $\bar{v}_{obs} \sim 40 \; \mathrm{km \; s^{-1}}$.

The rate of occultations per star is given by,

\begin{equation}
\begin{aligned}
\label{eq:5}
    & \dot{N}_{O, \star} 
    \approx
    \\
    & 2 \pi \int_{1 \; \mathrm{AU}}^{L_1} n_O \left[\left(\frac{\rho_{min}\sqrt{\lambda D_L / 2}}{R_O}\right)^{-3.3} - \left(\frac{R_{max}}{R_O}\right)^{-3.3}\right]
    \\
    & \left(\frac{\bar{v}_{obs} D_L R_S}{D_S}\right) \; \mathrm{d}D_L \; \; +
    \\
    & 2 \pi \int_{L_1}^{L_2} n_O \left[\left(\frac{R_S D_L}{D_S R_O}\right)^{-3.3} - \left(\frac{R_{max}}{R_O}\right)^{-3.3}\right]
    \\
    & \left(\frac{\bar{v}_{obs} D_L R_S}{D_S}\right) \; \mathrm{d}D_L \; \; ,
\end{aligned}
\end{equation}
where the limits of the integrals, $L_1$ and $L_2$, are defined as follows:

\begin{equation}
    L_1 =
  \begin{cases}
    {\frac{\rho_{min}^2 \lambda D_S^2}{2 R_S^2}}, & \text{$\frac{\rho_{min}^2 \lambda D_S^2}{2 R_S^2} \leq {\frac{R_{max} D_S}{R_S}}$ \; \; ,}\\
    {\frac{2 R_{max}^2}{\rho_{min}^2 \lambda}}, & \text{$\frac{\rho_{min}^2 \lambda D_S^2}{2 R_S^2} \geq {\frac{R_{max} D_S}{R_S}}$ \; \; ,}\\
  \end{cases}
\end{equation}

\begin{equation}
    L_2 =
  \begin{cases}
    {\frac{R_{max} D_S}{R_S}}, & \text{$\frac{\rho_{min}^2 \lambda D_S^2}{2 R_S^2} \leq {\frac{R_{max} D_S}{R_S}}$ \; \; ,}\\
    {0}, & \text{$\frac{\rho_{min}^2 \lambda D_S^2}{2 R_S^2} \geq {\frac{R_{max} D_S}{R_S}}$ \; \; .}\\
  \end{cases}
\end{equation}

\begin{figure*}[!th]
  \centering
  \includegraphics[width=.55\linewidth]{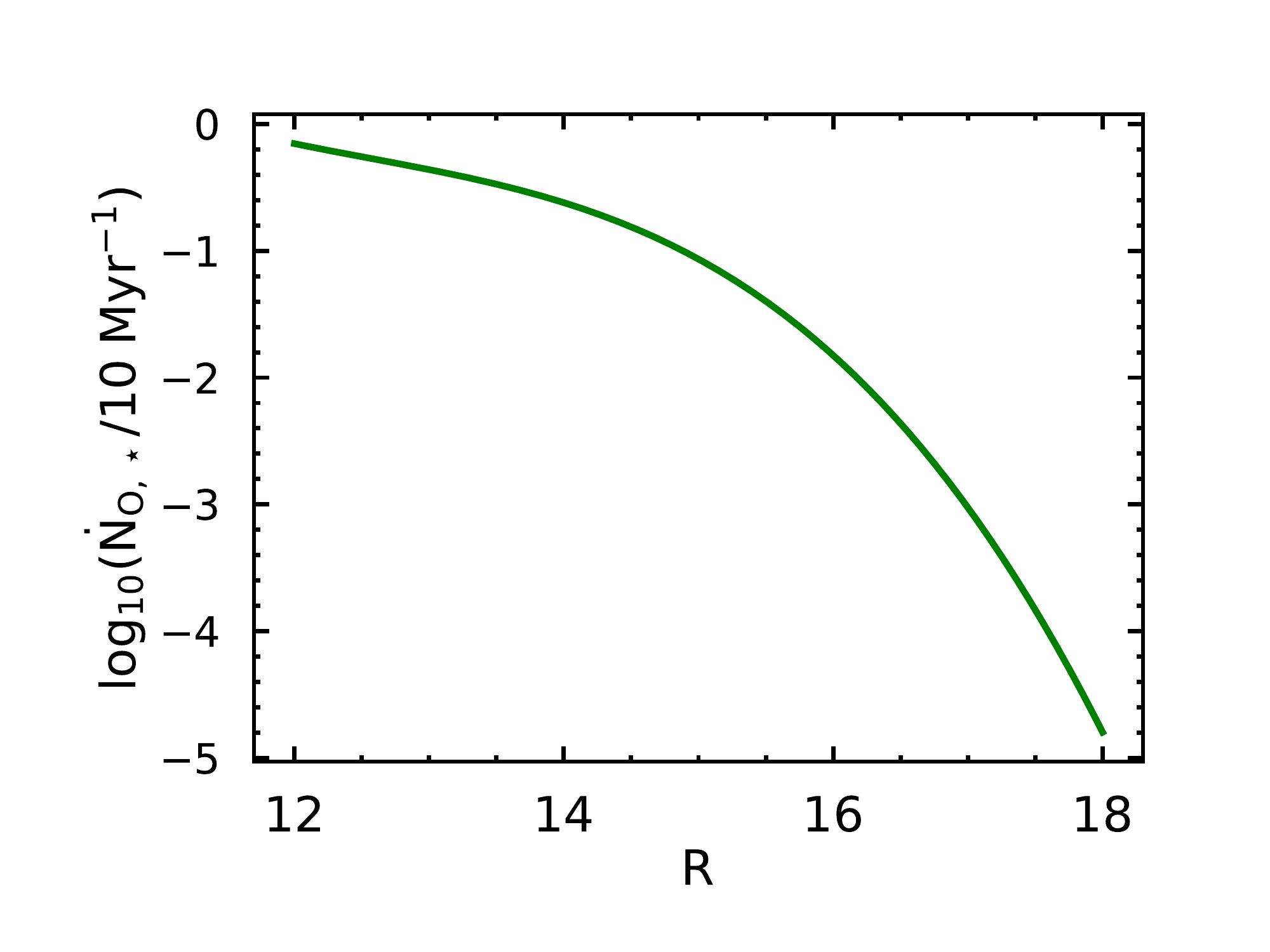}
    \caption{ISO occultation rate per star as a function of R magnitude for solar-type stars.}
    \label{fig:other}
\end{figure*}

\begin{figure*}[!th]
  \centering
  \includegraphics[width=.55\linewidth]{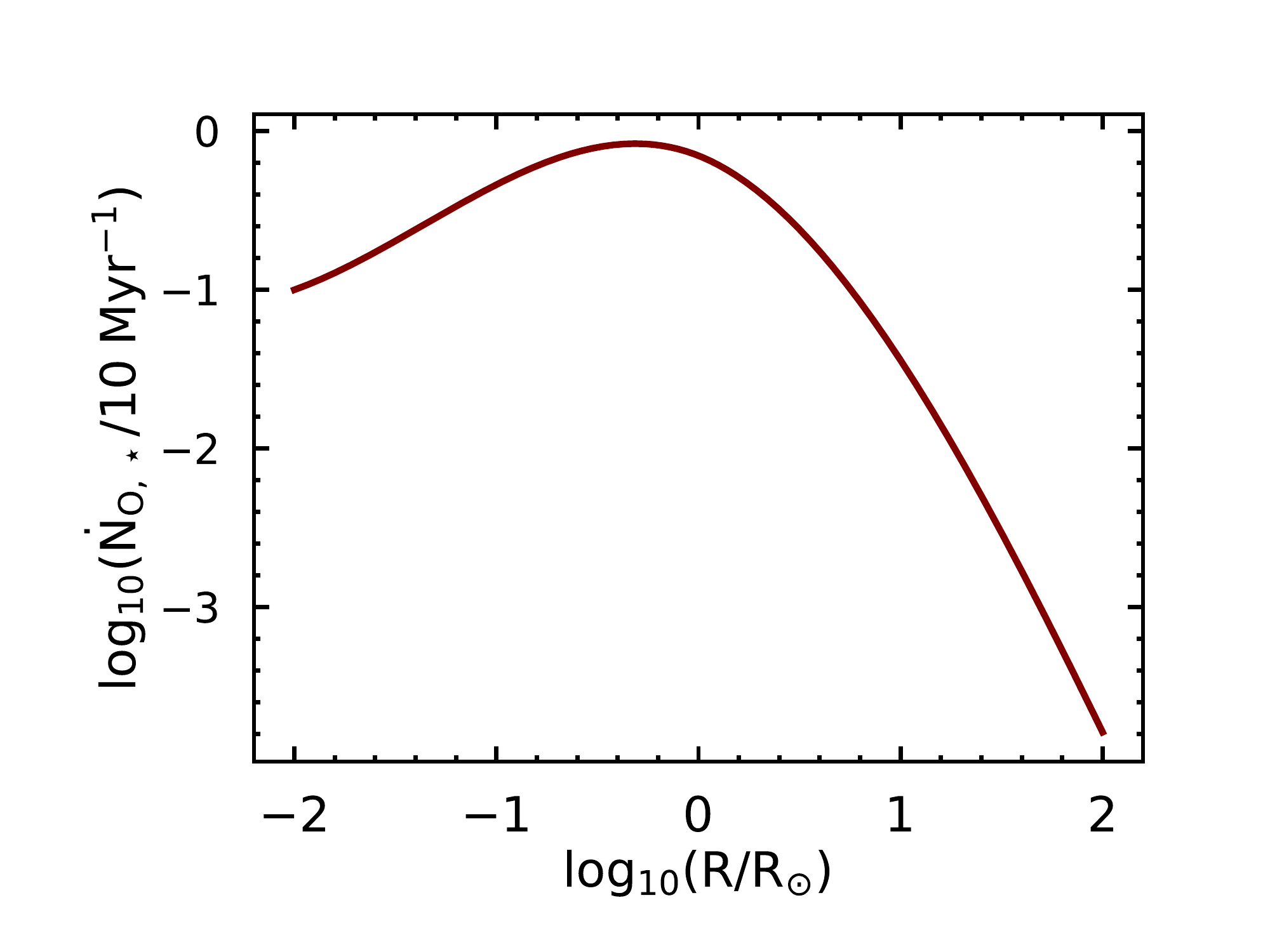}
    \caption{ISO occultation rate per star as a function of stellar radius, for a R = 12 star.}
    \label{fig:other}
\end{figure*}

\begin{figure*}[!th]
  \centering
  \includegraphics[width=.55\linewidth]{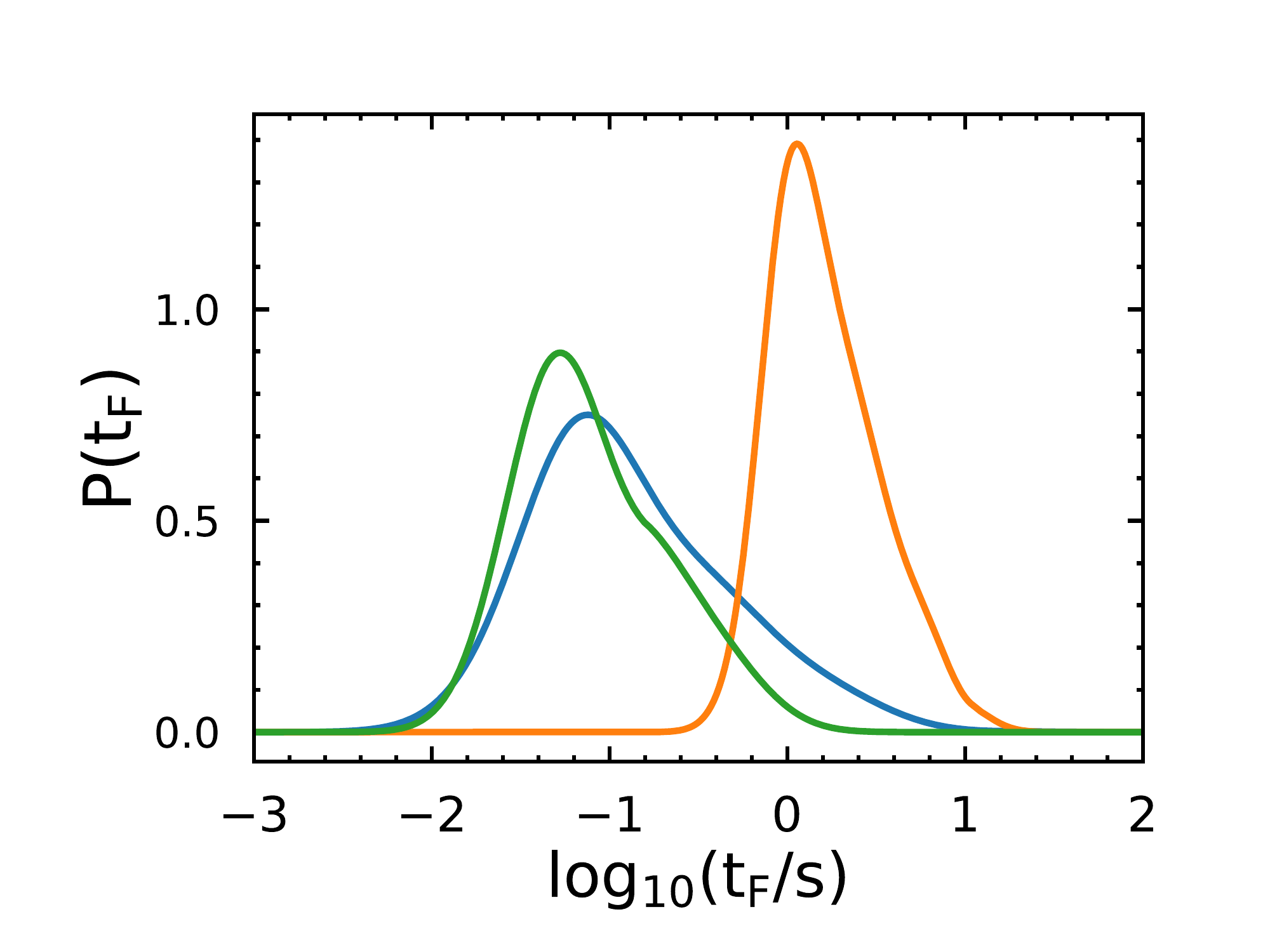}
    \caption{Expected probability distribution of Fresnel crossing times during stellar occultations observed from Earth for ISOs (blue), KBOs (green), and OCOs (orange), for a R = 12 solar-type star.}
    \label{fig:fig3}
\end{figure*}

The resulting ISO occultation rate per star as a function of magnitude is shown in Figure \ref{fig:other}, and corresponds to $\sim 0.1 \; \mathrm{Myr^{-1}}$ at R = 12.

To determine how easily discernable ISO, KBO, and OCO occultation signatures are from each other, and to understand the distribution of timescales on which ISO occultations occur, we numerically simulated the distribution of expected Fresnel distance crossing times for each population, assuming a R = 12 solar-type star.

We draw distances from the distribution $P(D_L) \propto D_L^2$ with bounds of 1 AU to $(R_{max} D_S / R_S)$ for ISOs, bounds of 30 to 50 AU for KBOs, and bounds of $2 \times 10^3$ to $10^5$ AU for OCOs. We draw velocities from the aforementioned kinematics for ISOs; we compute the orbital speed corresponding the distance for KBOs and for OCOs. We do not consider gravitational focusing and acceleration by the Sun for ISOs as these effects would only be significant at $D_L \lesssim 1\; \mathrm{AU}$. We sample random points in the Earth's orbit to obtain the motion of the observer relative to the object. We draw sizes for each population from a power law distribution with index $\sim -3$, with the appropriate bounds for each population. If conditions (\ref{eq:3}) and (\ref{eq:4}) are satisifed, we compute the Fresnel distance crossing time, $t_F = (F / v_{\perp})$. Otherwise, we re-draw distance and size. The resulting distributions of $t_F$ are shown in Figure \ref{fig:fig3}. ISO occultation events have have an even distribution with distance between 1 AU and $\sim 10^3$ AU. We find that the timescales for ISO and OCO occultation events are distinct. The ISO timescale distribution peaks at $t_F = 0.1 \; \mathrm{s}$, and a survey can avoid KBOs by pointing away from the plane of the ecliptic.

\section{Conclusions}

There are $\sim 7 \times 10^6$ stars in a with magnitude $R \lesssim 12.5$ \citep{2018RNAAS...2...51M}. An all-sky network of 1-m telescopes continuously monitoring all R = 12 stars with a time resolutions of $0.1 \; \mathrm{s}$ should yield a discovery rate of $\sim 1$ interstellar object per year. This would be a significant improvement on the current discovery rate of an ISO every few years. Our method supplements direct detection through reflected sunlight but nearby events could benefit from both methods of detection. By measuring each occultation in two or three colors, the radius and distance of the occulting object can be constrained \citep{Dyson1992}.

The data from such a survey would provide invaluable new information on the size distribution, composition, and possible origin of ISOs. Such information would be particularly valuable given the puzzle of the first two confirmed interstellar objects, `Oumuamua and Borisov, having such different physical characteristics \citep{Meech2017, Guzik2019}, and given the high implied abundance of ISOs relative to previous predictions \citep{Moro-Martin2009}.


\section*{Acknowledgements}
This work was supported in part by a grant from the Breakthrough Prize Foundation. 

\end{document}